\documentstyle[epsfig,12pt]{article}

\begin{document}

\title{{\large {\bf Anomaly Matching in Gauge Theories at Finite Matter Density}}}
\author{Stephen D.H.~Hsu\thanks{%
hsu@duende.uoregon.edu} \\
Department of Physics, \\
University of Oregon, Eugene OR 97403-5203 \\
\\
Francesco Sannino\thanks{%
francesco.sannino@yale.edu} \\
Department of Physics, \\
Yale University, New Haven, CT 06520 \\
\\
Myck Schwetz\thanks{%
ms@bu.edu} \\
Department of Physics,\\
Boston University, Boston, MA 02215, USA}
\date{June, 2000}
\maketitle

\begin{abstract}
We investigate the application of 't Hooft's anomaly matching conditions to
gauge theories at finite matter density. We show that the matching
conditions constrain the low-energy quasiparticle spectrum associated with
possible realizations of global symmetries.
\end{abstract}



\newpage

\section{Introduction}

Quark matter at high density provides an interesting laboratory for the
controlled study of non-perturbative physics \cite{quarkmatter}. Novel
phenomena such as color superconductivity and color-flavor locking have been
shown to result from physics near the quark Fermi surface. Rigorous results
can be obtained in the limit of asymptotic density. Less, however, is known
about what happens at intermediate densities, where the effective QCD
coupling is still large. The intermediate density region is more likely to
be realized in neutron star cores or laboratory experiments.

The 't Hooft anomaly matching conditions \cite{Hooft} constrain the
realizations of chiral symmetry in the low energy phase of a gauge theory.
In this paper we investigate whether these conditions, orginally derived at
zero density, can be used to constrain the behavior of matter at non-zero
density. There are several obstacles to this generalization. A basic
observation is that Lorentz symmetry, which is broken at non-zero matter
density, plays an important role in the original derivation of these
matching conditions \cite{Hooft,CG,FSBR}, and in the physics of the anomaly.
In gauge theories at zero density, unbroken global chiral symmetries imply
the existence of massless spin-$1/2$ states. As elaborated in \cite
{CG,FSBR,DZ}, these degrees of freedom are responsible for the IR
singularities associated with the anomaly. It is not at all obvious that the
same considerations apply to the quasiparticles comprising the low-energy
spectrum at finite density. Yet, as noted by Sannino \cite{Sannino}, the
various phases of color superconductivity do indeed satisfy the anomaly
matching conditions.

't Hooft's original argument for his anomaly matching conditions involves
the use of ``spectator'' fermions to cancel any anomalies which result from
the gauging of flavor symmetries. In the low energy limit of the theory, one
is left only with the massless states (composite or elementary) and the
spectator fermions. If this long-wavelength theory is to be consistent, the
anomaly generated by the spectators must be cancelled by that of the other
massless states. This argument can be repeated in the case of finite
density, with gapless quasiparticles playing the role of massless fermions
in the low-energy effective theory. However, due to the existence of a Fermi
surface there is a large degeneracy of zero energy states. Naively, one
might expect an enhancement of the anomaly prefactor by an amount
proportional to the area of the Fermi surface, which scales like $\mu^2$
(the square of the chemical potential). Clearly, we need to understand more
precisely how quasiparticles contribute to the anomaly.

We will show that the anomaly matching conditions (AMCs) continue to apply
at finite density, and constrain the possible quasiparticle spectrum \cite
{Sannino}. In essence, the singularity structure of the three-current
correlator implied by the anomaly must be reproduced by the effects of
quasiparticles in the finite density theory, thereby constraining their
flavor quantum numbers. Applying the Landau-Cutkowsky rules, we see that
only quasiparticles satisfying special kinematic conditions contribute to
the singularity -- there is no degeneracy factor due to the Fermi surface.

We focus our attention on cold but dense matter theories while a
discussion of the anomaly at finite temperature can be found in
Ref.~\cite{PTT}.

This letter is organized as follows. In section 2 we discuss the
computation of the anomaly at finite density and show that it is
unaffected by the presence of a chemical potential. Further, the
anomaly continues to imply the existence of IR singularities, even
at non-zero density. In section 3 we use the Landau-Cutkowsky rules
to show that these singularities (in the absence of spontaneous
symmetry breaking) require the existence of gapless quasiparticles
in the low energy spectrum.

\section{Anomaly at Finite Density}

In this section we show that the anomaly is unaffected by the presence of a
chemical potential, as are the implications for singularities of the three
current correlator. Heuristically, density effects are IR in nature whereas
the anomaly can be computed from the UV behavior of the theory. Points (A)
and (B) below follow simply from this observation, although (C) does not.

{\bf A.} The standard derivation of the anomaly from the point of view of UV
divergencies involves the careful treatment of the variation of the
fermionic partition function \cite{Fuji}
\begin{equation}  \label{Z}
Z~=~ \int d{\cal M}(\psi,\bar{\psi}) \, exp\Big[ i \int d^4x \, \bar{\psi} %
\Big(i\, D\!\!\!\!/\Big) \psi \Big]~,
\end{equation}
under the infinitesimal change of variables
\begin{eqnarray}  \label{chirot}
\psi &\longrightarrow& (1 + i\alpha(x) \gamma^5)\, \psi~,  \nonumber \\
\bar{\psi} &\longrightarrow& \bar{\psi}\, (1 + i\alpha(x) \gamma^5)~.
\end{eqnarray}
Although the action in the exponent of the integrand is invariant
under (\ref {chirot}), the fermionic measure $d{\cal M}$ is not:
$d{\cal M} \rightarrow {\cal J}^{-2}\,d {\cal M}$. The Jacobian
determinant ${\cal J}$ is evaluated using gauge invariant
regularization in the basis of the eigenstates of Dirac operator
$D\!\!\!\!/$, i.e. $\psi(x) = \sum_n a_n \phi_n(x)~, \bar{\psi }(x)
= \sum_n \tilde{a}_n \tilde{\phi}_n(x)$ and $d{\cal M} = \Pi da_m
d
\tilde{a}_m$.

Note, that if the chemical potential is not zero, one may still
define the measure in the basis of the eigenstates of Dirac
operator with zero chemical potential. All further considerations
follow the same path and produce the familiar result
\begin{equation}  \label{jacob}
{\cal J}~=~ exp\Big[ -i \int d^4x \, \alpha(x) \, \Big(\, {\frac{e^2 }{32
\pi^2}} \epsilon^{\mu\nu\rho\sigma} F_{\mu\nu}F_{\rho\sigma}(x) \,\Big)\,
\Big]
\end{equation}
This implies that the anomaly equation for the axial current will take the
same form at non-zero fermion density:
\begin{equation}  \label{axial}
\partial_\mu j_5^{\mu} ~=~ - {\frac{e^2 }{32 \pi^2}} \epsilon^{\mu\nu\rho
\sigma}F_{\mu\nu}F_{\rho\sigma}
\end{equation}

{\bf B.} It is known that the anomalies arise via quantum corrections, (i.e.
renormalization). In particular axial anomalies are associated, in a
diagrammatic language, with triangular diagrams.

The three point function in electrodynamics is: (the generalization
to non-Abelian theories is straightforward)
\begin{equation}  \label{T}
T_{\mu \nu \lambda}\left(k_1,k_2,q\right) = i\, \int\, d^4x_1\, d^4x_2
\left<0|T\left(V_{\mu}(x_1) V_{\nu}(x_2) A_{\lambda}(0) \right) |\,0\right>
\, e^{ik_1\cdot x_1 + i k_2\cdot x_2} \ ,
\end{equation}
with $q=k_1+k_2$ and $V$ and $A$ vector and axial currents
respectively. Since the anomaly is independent of a fermion mass,
we set it to be zero. The classical axial Ward identity
$q^{\lambda} T_{\mu \nu \lambda}=0$ is modified according to (for a
more detailed discussion see \cite{ChengLi} before and after
formulae (6.31-32)):
\begin{equation}
q^{\lambda} T_{\mu \nu \lambda}=\Delta_{\mu \nu}^{(1)} + \Delta_{\mu
\nu}^{(2)} \ ,
\end{equation}
with
\begin{equation}
\Delta_{\mu \nu}^{(1)}= \int \frac{d^4p}{(2\pi)^4} {\rm Tr} \left[\frac{i}{%
p\!\!\!/} \gamma_5 \gamma_{\nu} \frac{i} {p\!\!\!/ - k\!\!\!/_1 }
\gamma_{\mu} - \frac{i}{p\!\!\!/- k\!\!\!/_2} \gamma_5 \gamma_{\nu} \frac{i%
} {p\!\!\!/ - q\!\!\!/ } \gamma_{\mu}\right] \ ,  \label{d1}
\end{equation}
and $\Delta_{\mu \nu}^{(2)}$ is obtained by interchanging $\mu \rightarrow
\nu$ as well as $k_1 \rightarrow k_2$. Note that if we could shift the
integration variable $p$ to $p+k_2$ in the second term of (\ref{d1}) the $%
\Delta^{(1)}_{\mu \nu}$ term would vanish identically. Since the integrals
are linearly divergent, it can be shown that (at zero density) a translation
of the integration variable produces extra finite terms ruining the
classical Ward identity.

It is natural to ask what happens to the triangle anomaly at finite matter
density. In principle, the contributions from triangle diagrams could depend
on the chemical potential $\mu$: $\Delta^{(i)}_{\mu\nu}(\mu)$.

We compute the difference $\Delta^{(i)}_{\mu\nu}(\mu) -
\Delta^{(i)}_{\mu\nu}(0)$ parameterizing a possible deviation in the anomaly
calculations. We explicitly evaluate the finite density effects for the $i=1$
case (the effects for the $i=2$ are identical):
\begin{eqnarray}  \label{d}
\delta\Delta_{\mu\nu}^{(1)}(\mu) ~=~ \int {\frac{d^4p }{(2\pi)^4}} {\rm Tr}%
\left[ {\frac{i }{p\!\!\!/ - \mu \gamma_0}}\gamma_5 \gamma_{\nu} {\frac{i }{%
p\!\!\!/ - k\!\!\!/_1 - \mu \gamma_0}}\gamma_{\mu} ~-~ {\frac{i }{p\!\!\!/}}%
\gamma_5 \gamma_{\nu} {\frac{i }{p\!\!\!/ - k\!\!\!/_1}}\gamma_{\mu} -
\right.  \nonumber \\
\left. \Big( {\frac{i }{p\!\!\!/ - k\!\!\!/_2 - \mu \gamma_0}}\gamma_5
\gamma_{\nu} {\frac{i }{p\!\!\!/ - q\!\!\!/ - \mu \gamma_0}}\gamma_{\mu} ~-~
{\frac{i }{p\!\!\!/ - k\!\!\!/_2}}\gamma_5 \gamma_{\nu} {\frac{i }{p\!\!\!/
- q\!\!\!/}}\gamma_{\mu}\Big) \right] \ .
\end{eqnarray}
By applying a momentum shift ($p$ to $p-u$), with the shift vector $%
u^{\tau}=(\mu, 0, 0, 0)$, we expect a possible non null contribution to be
encoded in the following integral (see \cite{ChengLi}):

\begin{equation}
\delta{\Delta ^{(1)}_{\mu\nu}}(\mu)= u^{\tau} \int \frac{d^4p}{(2\pi)^4}
\frac{\partial}{\partial p^{\tau}} f(p,k_1,k_2)\ ,
\end{equation}
and
\begin{equation}
f(p,k_1,k_2)=- i\,4 \epsilon_{\rho\sigma\mu \nu} \left[\frac{%
p^{\rho}\left(p-k_1\right)^{\sigma}}{p^2 \left(p-k_1\right)^2} - \frac{%
\left(p-k_2\right)^{\rho}\left(p-q\right)^{\sigma}} {(p-k_2)^2
\left(p-q\right)^2}\right] \ .
\end{equation}
Applying Gauss's theorem for the case of four-dimensional Minkowski space,
we have:
\begin{eqnarray}
\delta{\Delta ^{(1)}_{\mu\nu}}(\mu)&=& u^{\tau} \int \frac{d^4p}{(2\pi)^4}
\frac{\partial}{\partial p^{\tau}} f(p,k_1,k_2) = u^{\tau}\frac{2i\pi^2}{%
(2\pi)^4}\lim_{p\rightarrow \infty}\, p^2 p_{\tau} f(p,k_1,k_2)  \nonumber \\
&=& \frac{\mu^{\tau}}{2\pi^2}\, \epsilon_{\rho \sigma \mu \nu} \lim_{p
\rightarrow \infty} p_{\tau} \left(- \frac{p^{\rho}k_1^{\sigma}}{p^2} ~+~
\frac{p^{\rho}q^{\sigma} + p^{\sigma} k_2^{\rho} }{p^2}\right) \\
&=& \frac{\epsilon_{\rho \sigma \mu \nu}}{8\pi^2} \Big( u^{\rho}
k_2^{\sigma} ~+~ u^{\sigma} k_2^{\rho}\Big)~=~ 0 \ ,  \nonumber
\end{eqnarray}
and, in the last step, we used $\displaystyle{\lim_{p \rightarrow \infty}%
\frac{p^{\tau}p^{\rho}}{p^2}=\frac{g^{\tau \rho}}{4}}$.

Thus, for any finite density we have no extra contribution to the axial
anomaly with the respect to the zero density one.

Adler and Bardeen \cite{AB} showed, at zero density, that the axial anomaly
coefficient is one loop exact. We expect this theorem to hold at finite
density, although we will not attempt to provide a proof. Heurisitcally we
can say that the introduction of a density term does not affect the
superficial degree of divergence of the higher-order triangle diagrams with
respect to the zero density case. Since the latter have a lower degree of
divergence with the respect to the simple triangle diagram we have no
momentum-routing ambiguity and hence vanishing contribution to the anomaly.

{\bf C.} At zero density the anomaly implies an IR singularity in
the three-point function (\ref{T}) involving vector and axial
currents \cite{CG}.

Now the form of $T_{\mu\nu\lambda}$ is strongly restricted by
Lorentz, Bose and permutation symmetries:

\begin{eqnarray}
T_{\sigma \rho \mu}\left(k_1,k_2,q\right)=&+&
F_1\left(k_1^{\tau}-k_2^{\tau} \right) \epsilon_{\tau \sigma \rho
\mu} + F_2 \left[k_{1\sigma} k_1^{\alpha}k_2^{\beta}
\epsilon_{\alpha\beta\rho\mu} -k_{2\rho} k_1^{\alpha}k_2^{\beta}
\epsilon_{\alpha\beta\sigma\mu} \right] \nonumber \\
&+& F_3 \left[k_{2\sigma} k_1^{\alpha}k_2^{\beta}
\epsilon_{\alpha\beta\rho\mu} -k_{1\rho} k_1^{\alpha}k_2^{\beta}
\epsilon_{\alpha\beta\sigma\mu} \right] + \dots \ .
 \label{TCL}
\end{eqnarray}
 In general (\ref{TCL}) may contain tensor-like structures
(denoted here by $
\dots$) which unlike pseudo-tensorial ones do not contribute to the anomaly.

At non-zero matter density $T_{\mu\nu\lambda}$ contains more terms,
due to the presence of a constant Lorentz four-vector $u = (\mu,
\vec{0})$. These terms will, however, be still restricted by the
rotational $SO(3) $, Bose and permutation symmetries. It is
convenient to introduce the 4-vector $\eta$  with $u=\mu \eta$. Due
to the presence of a new independent 4-vector $\eta$ the $F$
functions, are general functions, i.e. $F=F(k_1\cdot
\eta, k_2\cdot
\eta, \mu, k_1^2,k_2^2)$. Following Ref.~\cite{PTT} the most
general decomposition in terms of invariant amplitudes is:
\begin{eqnarray}
T_{\sigma \rho \mu}\left(k_1,k_2,q\right)&=& +
F_1\left(k_1^{\tau}-k_2^{\tau} \right) \epsilon_{\tau \sigma \rho
\mu} + F_2 \left[k_{1\sigma} k_1^{\alpha}k_2^{\beta}
\epsilon_{\alpha\beta\rho\mu} -k_{2\rho} k_1^{\alpha}k_2^{\beta}
\epsilon_{\alpha\beta\sigma\mu} \right] \nonumber \\
&+& F_3 \left[k_{2\sigma} k_1^{\alpha}k_2^{\beta}
\epsilon_{\alpha\beta\rho\mu} -k_{1\rho} k_1^{\alpha}k_2^{\beta}
\epsilon_{\alpha\beta\sigma\mu} \right] + F_4 \left[\eta_{\sigma} k_1^{\alpha}k_2^{\beta}
\epsilon_{\alpha\beta\rho\mu} -\eta_{\rho} k_1^{\alpha}k_2^{\beta}
\epsilon_{\alpha\beta\sigma\mu} \right] \nonumber \\
&+& F_5 \left[k_{1\sigma} k_1^{\alpha}\eta^{\beta}
\epsilon_{\alpha\beta\rho\mu} + k_{2\rho} k_2^{\alpha}\eta^{\beta}
\epsilon_{\alpha\beta\sigma\mu} \right] + F_6 \left[k_{2\sigma} k_1^{\alpha}\eta^{\beta}
\epsilon_{\alpha\beta\rho\mu} +k_{1\rho} k_2^{\alpha}\eta^{\beta}
\epsilon_{\alpha\beta\sigma\mu} \right] \nonumber \\
&+& F_7 \left[\eta_{\sigma} k_1^{\alpha}\eta^{\beta}
\epsilon_{\alpha\beta\rho\mu} + \eta_{\rho} k_2^{\alpha}\eta^{\beta}
\epsilon_{\alpha\beta\sigma\mu} \right] + F_8\left[k_{1\rho} k_1^{\alpha}\eta^{\beta}
\epsilon_{\alpha\beta\sigma\mu} +k_{2\sigma} k_2^{\alpha}\eta^{\beta}
\epsilon_{\alpha\beta\rho\mu} \right] \nonumber \\
&+& F_9 \left[k_{2\rho} k_1^{\alpha}\eta^{\beta}
\epsilon_{\alpha\beta\sigma\mu} + k_{1\sigma} k_2^{\alpha}\eta^{\beta}
\epsilon_{\alpha\beta\rho\mu} \right] + F_{10} \left[\eta_{\rho} k_1^{\alpha}\eta^{\beta}
\epsilon_{\alpha\beta\sigma\mu} +\eta_{\sigma} k_2^{\alpha}\eta^{\beta}
\epsilon_{\alpha\beta\rho\mu} \right] \nonumber \\
&+&\left[F_{11}\left(k_{1\rho} k_{1\sigma} - k_{2\rho}
k_{2\sigma}\right) +F_{12}\left(k_{1\rho} \eta_{\sigma} -
\eta_{\rho} k_{2\sigma}\right) \right. \nonumber \\  &&\left. +
F_{13}\left(k_{2\rho} \eta_{\sigma} - \eta_{\rho}
k_{1\sigma}\right)
\right] k_1^{\alpha} k_2^{\beta} \eta^{\gamma}
\epsilon_{\alpha\beta\gamma\mu} \ .
\end{eqnarray}
One can explicitly verify that the previous amplitude is even under
the simultaneous exchange of $\rho \rightarrow \sigma$ and $k_1
\rightarrow k_2$. In the previous expression we dropped the flavor structure of the three point
function since it is not relevant for our discussion.

In what follows
we fix the kinematics such that $k_1^2=k_2^2=k^2$ with $k_1\cdot \eta=k_2\cdot \eta= k_0$.

Further constraints arise when imposing the anomaly equation along with the conservation of the vector currents (i.e. in formulae):
\begin{eqnarray}
q^{\mu} T_{\sigma \rho \mu} \left(k_1,k_2,q\right) &=&
\frac{i}{2\pi^2} k_1^\alpha k_2^{\beta} \epsilon_{\alpha\beta\sigma\rho} \ , \\
k_1^{\sigma} T_{\sigma \rho \mu} \left(k_1,k_2,q\right) &=&
k_2^{\rho} T_{\sigma \rho \mu} \left(k_1,k_2,q\right) = 0 \ .
\label{newcon}
\end{eqnarray}

After imposing these constraints and conveniently dotting the final
results in the vectors $\eta$, $k_1$ and $k_2$ (see
Ref.~\cite{PTT}) we have the following relations:
\begin{eqnarray}
2F_1+\left(F_5+F_6-F_8-F_9\right) k_0 +
\left(F_7-F_{10}\right)&=&-\frac{i}{2\pi^2} \ , \nonumber \\
\left(F_5-F_9\right)k^2 + \left(F_6-F_8\right)k_1\cdot k_2 + \left(F_7 - F_{10} \right)k_0&=& 0 \ ,
\nonumber  \\
\left(F_5-F_9\right)k_1\cdot k_2 + \left(F_6-F_8\right)k^2 + \left(F_7 - F_{10} \right)k_0&=& 0 \ .
\end{eqnarray}
The latter equations are due to the anomaly condition in Eq.~(\ref{newcon}).
The vector current conservation, on the other side, leads to:
\begin{eqnarray}
F_1+F_3k_1\cdot k_2+ F_2 k^2+F_4 k_0  &=&0 \ , \nonumber \\ F_6
k_1\cdot k_2 + F_5 k^2 + F_7 k_0  &=& 0 \ , \nonumber \\ F_8
k_1\cdot k_2 + F_9 k^2 + F_{10} k_0  &=& 0 \ .
\end{eqnarray}
Combining the previous relations we have the following anomaly constraint:
\begin{equation}
2F_3 k_1\cdot k_2 + 2k^2 F_2 + 2 k_0  F_4 -
\frac{1}{2}\frac{{\vec{q}\cdot \vec{q}}}{k_0} \hat{F}=\frac{i}{2\pi^2} \ ,
\end{equation}
with $\hat{F}=F_5 - F_9 = F_6 - F_8$.

In the absence of Lorentz breaking (i.e. zero density) $F_4=\hat{F}=0$ and a singularty in $q^2=(k_1 + k_2)^2 \rightarrow 0$ arises when considering the limit $k_1^2 =k_2^2=k^2=0$:
\begin{equation}
\lim_{k^2\rightarrow 0} F_3 = \frac{i}{2\pi^2 q^2} \ .
\end{equation}
This pole is interpreted as due to the pion when chiral symmetry is
spontaneously broken.

What happens at finite density ? Following Ref.~\cite{PTT} let us define
the following null vectors $k_1=k_0(1,0,0,1)$ and $k_2=k_0(1,\sin\theta,0,
\cos\theta)$. In these variables we have:
\begin{equation}
F_3=\frac{i}{4\pi^2 k_0^2\left(1-\cos\theta\right)} -
\frac{F_4}{k_0\left(1-\cos\theta\right)}+
\frac{\hat{F}\left(1+\cos\theta\right)}{\left(1-\cos\theta\right)} \ .
\end{equation}
Let $\cos\theta \ne 1$, and consider the double limit
$k_0\rightarrow 0$ and $q^2 \rightarrow 0$.
We deduce ($q^2 = 2 k_0^2 (1 - \cos \theta)$):
\begin{equation}
\lim_{k^2\rightarrow 0;q_0\rightarrow 0}F_3=\frac{i}{2\pi^2 q^2} \ .
\end{equation}
The form of this singularity is the same as the zero density
singularity, as long as we consider the $k_0 \rightarrow 0$ limit.
We again interpret this singularity as due to massless excitations.
In the case of unbroken chiral symmetry it implies the existence of
gapless excitations with the charges necessary to match the zero
density anomaly.

It is important to note that this singularity does {\it not} imply
that the anomalous decay $\pi \rightarrow 2 \gamma$ is unaffected
by finite density effects (see \cite{PTT}). In medium, the gauge
boson acquires a plasma mass, and hence the physical decay takes
place at non-zero energy. At such kinematic points we cannot
isolate the singularity cleanly, and density effects are possible.

In order to better clarify how the infrared singularities emerge at
finite density, we investigate the singularity properties of
quasiparticle scattering
in the following section.

\section{Quasiparticles and Singularities}

Since the anomaly equation (\ref{axial}) is unmodified by finite density
effects, we can apply 't Hooft's spectator fermion construction to conclude
that quasiparticles must generate the same anomalies as the fundamental
fermions. In this section we examine in more detail how this occurs. We are
particularly interested in why the Fermi surface degeneracy does not alter
the result, and why massive excitations do not contribute to the anomaly.

As discussed above, the anomaly at finite density still implies a $1/Q^2$
pole in the function $F_1 (Q^2)$. We can use the Landau-Cutkowsky rules \cite
{LC}, to identify the intermediate states which are capable of providing
this singularity. We first prove that the singularity structure is not
modified by matter density effects. Then we investigate the kinematics
associated with the singularity to gain a better understanding of why this
is the case.

Using the Landau-Cutkowsky rules \cite{LC}, the discontinuity in the
triangle graph is given by:
\begin{eqnarray}  \label{discT}
{\rm Disc}~ T_{\mu \nu \lambda} &\propto& \int \frac{dp_0 \, d^3p}{(2\pi)}
\left(p+k_1-u\right)_{\rho}\left(p+k_1+k_2-u\right)_{\sigma}
\left(p-u\right)_{\tau} \Theta\left(p^0-\mu\right) \delta\left[%
\left(p-u\right)^2\right]\,  \nonumber \\
&~&\Theta\left(p^0+k_1^0-\mu\right) \delta\left[\left(p+k_1-u\right)^2\right]
\, \Theta\left(p^0+k_1^0+k_2^0-\mu\right) \delta\left[\left(p+k_1+k_2-u%
\right)^2\right],  \nonumber \\
\end{eqnarray}
times the trace factor ${\rm Tr}\left[\gamma_5
\gamma^{\mu}\gamma^{\rho}\gamma^{\nu}\gamma^{\sigma}\gamma^{\lambda}\gamma^{%
\tau}\right]$. Here we used the result \cite{LC} that the leading
singularity in the physical region is given by the graph(s) in which each
internal particle is on-shell and propagates forward in time. In other
words, we replaced each internal propagator by $2 \pi i \Theta(l_0-\mu)
\delta\left[(l_0-\mu)^2 -\vec{l}^2 \right]$, where $l$ is the particle
momentum.

We can now perform the integral in $p_0$ by using the first $\delta$
function, which requires that $p_0 - \mu = \vert \vec{p} \vert$. We obtain
\begin{eqnarray}
{\rm Disc}~ T_{\mu \nu \lambda} &\propto&\int \frac{d^3p}{2|\vec{p}|(2\pi)}
\left(\tilde{p}+k_1\right)_{\rho}\left(\tilde{p}+k_1+k_2\right)_{\sigma}
\left(\tilde{p}\right)_{\tau}\,  \nonumber \\
&~&\Theta \left( |\vec{p}|+k_1^0 \right) \delta\left[\left(\tilde{p}
+k_1\right)^2\right] \, \Theta\left(|\vec{p}|+k_1^0+k_2^0\right) \delta\left[
\left(\tilde{p}+k_1+k_2\right)^2\right] \ ,
\end{eqnarray}
with the 4-momentum $\tilde{p}=\left(|\vec{p}|,\vec{p}\right)$.
This integral no longer has any explicit dependence on the finite
density term $\mu$, and is identical to the integral obtained in
the zero density case. We see that $p_0 - \mu$ in the finite
density case merely plays the role of the energy $p_0$ in the zero
density case. Hence the singularity structure is unchanged at
finite density.

\begin{figure}[htb]
\center{ \epsfysize=8.0 cm \leavevmode \epsfbox{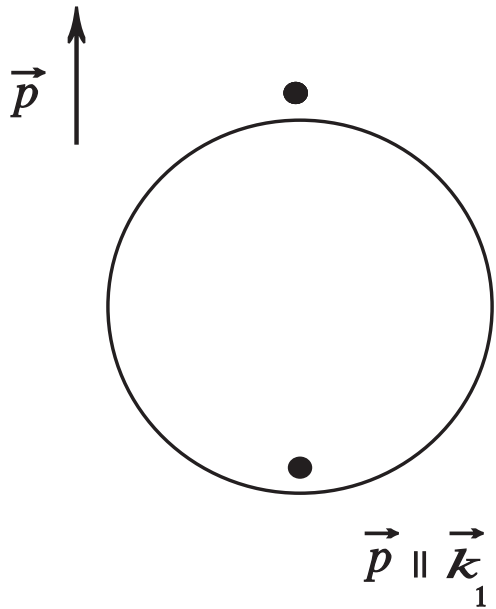}
\caption{The particle-hole pair contributing to the anomaly singularity
is determined by the external momentum $k_1$.}\label{fermi}}
\end{figure}

Finally, we would like to understand why the Fermi surface degeneracy does
not affect the anomaly calculation. To this end, let us examine the on-shell
kinematic conditions satisfied by the quasiparticles in (\ref{discT}).
Suppose we regard the triangle graph as a spacetime process, with the $k_1$
vertex earliest in time. Let the incoming gauge boson momentum $k_1$ be
null. In order that the quasiparticles with momenta $p$ and $p + k_1$ both
be on-shell, $\vec{p}$ and $\vec{k}_1$ must be parallel. Thus, a particular
point on the Fermi surface is selected by $\vec{k_1}$: the rest of the
surface cannot contribute to the singularity (see figure \ref{fermi}). (One
of the particles emerging from the $k_1$ vertex is actually a quasihole on
the opposite side of the Fermi surface from the quasiparticle. Which is
which depends on whether $\vec{k}_1$ and $\vec{p}$ are aligned or
anti-aligned.) Finally, we note that if we take the energy of the gauge
boson $k_1^0 \rightarrow 0$, we will not be able to produce quasiparticles
whose energy spectrum has a gap: only gapless quasiparticles can reproduce
the anomaly in the entire physical region.

\section{Discussion}

As shown, 't Hooft's anomaly matching conditions are as valid at non-zero
matter density as at zero density. They provide a non-trivial check on
recent results on QCD at asymptotic density. Morever, they apply even at
strong coupling, where dynamical calculations cannot be reliably performed.

\bigskip \noindent {\bf Acknowledgements}

\noindent The authors would like to thank Robert Pisarski, Krishna Rajagopal and
Thomas Sch\"{a}fer for useful discussions and comments. S.H. and
M.S. thank the Institute for Nuclear Theory at the University of
Washington, where this work was begun, for its hospitality. F.S.
would also like to thank J.~Schechter and Z.~Duan for helpful
discussions. This work was supported in part under DOE contracts
DE-FG02-91ER40676, DE-FG-02-92ER-40704 and DE-FG06-85ER40224.

\end{document}